# Soft Computing Techniques in combating the complexity of the atmosphere- a review

Surajit Chattopadhyay**

1/19 Dover Place

Kolkata 700 019

West Bengal

India

**E-mail      srajit_2008@yahoo.co.in

## Abstract

The purpose of the present review is to discuss the role of Soft Computing techniques in understanding the complexity associated with atmospheric phenomena and thus developing predictive models. Problems in atmospheric data analysis are discussed in brief and the relevance of Soft Computing to the atmospheric data analysis and their advantage over the conventional methods are also conversed. Applicability of different Soft Computing techniques is precisely discussed. In the last section, up-to-date literature appraisal is incorporated.

## 1.    Introduction

The Soft Computing techniques are based on the information processing in biological systems. The complex biological information processing system enables the human beings to survive with accomplishing tasks like recognition of surrounding, making prediction, planning, and acting accordingly. Human type information processing involves both logical and intuitive information processing. Conventional computer systems are good for the former, but their capability for the later is far behind that of human beings. For a computing system to have human like information processing facility, it should be flexible enough to support three features: openness, robustness, and real time processing. Openness of a system is its ability to adapt or extend on its own to cope with changes encountered in the real world. Robustness of a system means its stability and tolerability when confronted with distorted, incomplete, or imprecise information. The real time characteristic implies the ability of the system to react within a reasonable time in response to an event. Information processing systems with all these three characteristics are known as real world computing (RWC) systems (for details see Chaudhuri and Chattopadhyay, 2005a). A RWC system should, therefore, be capable of distributed representation of information, massively parallel processing, learning, and self-organization in order to achieve flexibility in information processing. Thus, Soft Computing can be viewed as the key ingredient of RWC systems. Several authors (Abraham e*t al*., 2001; Zhang and Scofield, 1994; Zhang and Knoll, 2001; Chaudhuri and Chattopadhyay, 2005a,b) have discussed the potentials of the soft computing techniques in solving real world problems. Soft Computing has three basic components, namely, Artificial Neural Network, Fuzzy Logic, and Genetic Algorithm. In recent times, two other methods have been incorporated. They are Rough Set Theory and Ampliative Reasoning.

Atmosphere is a highly complex, dynamic, and non-linear system. It involves a large number of variables and phenomena. The interrelationships between the variables are mostly non-linear and most of the time series pertaining to different phenomena exhibit chaotic features. Research on the investigation of the existence of chaos in atmospheric processes has gained considerable attention in recent times (Handerson and Wells, 1988; Sharifi et al., 1990; Stehlik, 1999;Sivakumar et al, 1998;Sivakumar, 2000).

Problem in atmospheric data analysis and prediction arises mainly because of the intrinsic non-linearity and chaos (Sivakumar et al, 1998;Sivakumar, 2000). Principle et al (1992) implemented Artificial Neural Network in predicting chaotic time series. Kozma et al (1998) discussed the potential of Soft Computing techniques in the form of Fuzzy Logic and Artificial Neural Network in predicting chaotic time series.

Artificial Neural Network (ANN) is useful in the situations, where underlying processes / relationships may display chaotic properties. ANN does not require any prior knowledge of the system under consideration and are well suited to model dynamic systems on a real-time basis (Maqsood *et al.*, 2002).

Fuzzy Logic is another Soft Computing technique necessary for analyzing complex systems, especially where the data structure is characterized by several linguistic parameters. Atmosphere being a highly complex system and various linguistic variables such as "hot", "cloudy", "severe" etc being highly relevant to it, fuzzy logic is highly suitable for analyzing the datasets pertaining to the atmospheric variables or occurrences.

Another component of soft computing techniques is genetic algorithm. This algorithm is based upon the concept of evolution and "survival of the fittest".

Soft Computing techniques, in recent times, have emerged as a convincing alternative to traditional procedures in analysis and prediction of weather phenomena over the globe. This sophisticated mathematical device has opened up new avenues to successful atmospheric research especially involving perilous weather events like thunderstorms, hailstorms, excess rainfall, flood etc. Soft Computing techniques are essentially suitable for atmospheric analysis and prediction because of its high flexibility, less data dependency, less dependency upon initial conditions, ability to work upon linguistic data, and ability to contend with chaotic features. All the components have individual potential and they can be summed up in hybrid Soft Computing models where more than one technique are mingled as complementary to each other. Artificial Neural Net can predict chaotic atmospheric time series and can recognize various complicated patterns intrinsic in the time series. Unlike numerical procedures, Artificial Neural Network requires no gridded structure and requires no prior knowledge regarding the distribution of the dataset at hand. This method being highly flexible, can handle any kind of chaotic change in the relevant input data. The most important usefulness of Artificial Neural Net over numerical weather procedures is its unresponsiveness to the initial conditions.

## 2.     Current status of research

Hu (1964) initiated the implementation of Artificial Neural Network, an important Soft Computing methodology in weather forecasting.  McCann (1992) developed Artificial Neural Network models to give 3-7 hr forecast of significant thunderstorms on the basis of surface based lifted index and surface moisture convergence. The two neural networks produced by them were combined operationally at National Severe Storms Forecast Center, Kansas City, Missouri to produce a single hourly product and was found to enhance the pattern recognition skill. Cook and Wolfe (1991) developed a neural network to predict average air temperatures. Zhang and Scofield (1994) applied Artificial Neural Network (ANN) in prediction of rainfall through satellite data. Allen and Le Marshall (1994) compared the performance of Artificial Neural Network approach and discrimination analysis method for operational forecasting of rainfall and established the superiority of neural network approach over conventional statistical approach in forecasting rainfall over Australia. Bankert and Aha (1996) adopted neural network approach for classification of cloud. Han and Felker (1997) implemented an ANN to estimate daily soil water evaporation from average relative air humidity, air temperature, wind speed, and soil water content in a cactus field study. They concluded that the ANN technique appeared to be an improvement over the multi-linear regression technique for estimating soil evaporation. Mohandes et al (1998) applied Artificial Neural Network for prediction of wind speed. Gardner and Dorling (1998) discussed the proficiency of Multilayer Perceptron as a suitable model for atmospheric prediction. Lee et al (1998) applied Artificial Neural Network in rainfall prediction by splitting the available data into homogeneous subpopulations. Wong et al (1999) constructed fuzzy rule bases with the aid of SOM and backpropagation neural networks and then with the help of the rule base developed predictive model for rainfall over Switzerland using spatial interpolation. Bruton et al. (2000) developed ANN models for estimating daily pan evaporation. The results were compared with those of multiple linear regression and Priestly-Taylor model and they found that the ANN model provided the highest accuracy. Perez et al (2000) applied Artificial Neural Network in prediction of atmospheric pollution. Jagdesh (2000) compared the performance of Artificial Neural Network with other empirical methods in prediction of watershed runoff. Hsieh and Tang (2001) elucidated the relevance of Artificial Neural Network in atmospheric and oceanic modeling. Perez and Reyes (2001) applied Multilayer Perceptron model in predicting the particulate air pollution. Li (2002) established the suitability of Artificial Neural Network model in estimating maximum surface temperature, minimum surface temperature, and solar radiation over regression method at Tifton, Georgia and Griffin. Maqsood *et al*. (2002) established the usefulness of Artificial Neural Network in atmospheric modeling explained its potential over conventional weather prediction model. Chaloulakou et al (2003) made a comparative study between multiple regression models and feed forward Artificial Neural Network with respect to their predictive potential for $PM_{10}$ over Athens and established that neural network approach has an edge over regression models expressed both in terms of prediction error and of episodic prediction ability. Rashidi and Rashidi (2004) developed an Artificial Neural Network in the form of Multilayer Perceptron to predict the solar activity, which is noticeably important for forecasting space weather. Chaudhuri and Chattopadhyay (2005a) developed an Artificial Neural Network model for prediction of some surface parameters during pre-monsoon thunderstorms over northeastern part of India. Jef et al (2005) developed an Artificial Neural Network model to predict daily average $PM_{10}$ concentration over Belgium.

Fuzzy Logic, another Soft Computing methodology, can also be of great use in atmospheric data analysis and prediction. Being capable of dealing with linguistic variables, this Soft Computing methodology can be of utilized in analyzing atmospheric variables. Bardossy et al (1995) implemented fuzzy logic in classifying atmospheric circulation patterns. Özelkan and Duckstein (1996) compared the performance of regression analysis and fuzzy logic in studying the relationship between monthly atmospheric circulation patterns and precipitation. Pesti et al (1996) implemented fuzzy logic in draught assessment. Baum et al (1997) developed cloud classification model using fuzzy logic. Lakshmanan and Witt (1997) implemented Fuzzy Logic in detecting severe updrafts. Fujibe (1998) classified the pattern of precipitation at Honshu with fuzzy

C-means method. Galambosi et al (1999) investigated the effect of ENSO and macro circulation patterns on precipitation at Arizona using Fuzzy Logic. Vivekanandan et al (1999) developed and implemented a fuzzy logic algorithm for hydrometeor particle identification that is simple and efficient enough to run in real time for operational use. Hansen (2000) applied fuzzy k-nn weather prediction system to improve the technique of persistence climatology by achieving direct, efficient, and expert-like comparison of past and present weather cases. Shao (2000) established fuzzy membership functions, based on cloud amount, cloud type, wind speed, and relative humidity, to compose a fuzzy function of weather categorization for thermal mapping. Liu and Chandrasekar (2000) developed a fuzzy logic and neuro-fuzzy system for classification of hydrometeor type based on polarimetric radar measurements, where fuzzy logic was used to infer hydrometeor type, and the neural network-learning algorithm was used for automatic adjustment of the parameters of the fuzzy sets in the fuzzy logic system according to prior knowledge. Gomez and Casanovas (2002) developed physical fuzzy model for solar irradiance by considering uncertainty associated with solar irradiance as fuzzy uncertainty and established the superiority of fuzzy model over conventional non-fuzzy models. Gautam and Kaushika (2002) implemented fuzzy logic in estimating the global solar radiation by considering cloudiness index, type of cloud, climatic region, season of most precipitation, and type of precipitation as the fuzzy random variables. They simulated their model for Calcutta and Delhi for the months of January and July. Bianco and Wilczak (2002) applied fuzzy logic to calculate the depth of the convective boundary layer, using vertical profiles of both radar-derived signal-to-noise ratio and variance of vertical velocity.

Genetic Algorithm is the component of Soft Computing that mainly deals with evolutionary processes on the basis of the concept of "survival of the fittest". Cartwright and Harris (1993) analyzed the distribution of airborne pollutants using Genetic Algorithm. Mulligan and Brown (1998) used Genetic Algorithm in calibrating water quality models. The bounded weak echo region (BWER) detection algorithm was developed by Lakshmanan (2000) using a genetic algorithm to tune fuzzy sets. Sen and Oztopal (2001) applied Genetic Algorithm in classification and prediction of precipitation occurrence. Haupt (2003) implemented Genetic Algorithm in geophysical fluid dynamics. Kishtawal et al (2003) assessed the feasibility of a nonlinear technique based on genetic algorithm for the prediction of summer rainfall over India. Bhattacharya (2004) applied Genetic Algorithm in combination with probability distribution in optimal design of unit hydrographs corresponding to two storm events over the catchment of North Potomac River of Maryland, USA. Yip and Wong (2004) applied Genetic Algorithm in fixing the tropical cyclone eye. They found the equations that best describe the temporal variations of the seasonal rainfall over India. Cheng et al (2005) made rainfall-runoff model calibration using a combination of parallel Genetic Algorithm and Fuzzy Logic. Quantitative ranges of some meteorological parameters associated with severe pre-monsoon thunderstorms were fixed by Chaudhuri and Chattopadhyay (2005b) using the theory of Genetic Algorithm. Kishtawal et al (2005) developed an automatic method for intensity estimation of tropical cyclones using multi-channel observations from TRMM Microwave Imager (TMI) using genetic algorithm. Peters et al (2003) applied Rough Set theory in classifying meteorological volumetric radar data used to detect storm events responsible for summer severe weather.

## 3. Conclusion

From the above review it is evident that Soft Computing techniques in various forms are moving forward as an impending tool to model a highly intricate system like atmosphere. More study on the chaotic feature of different atmospheric processes would reveal many new things when mingled with Soft Computing.

## References

Abraham, A., Philip, N.S., and Joseph, B., 2001, "Will we have a wet summer? Long term rain forecasting using Soft Computing models", *Modelling and Simulation 2001, Publication of the Society for Computer Simulation International, Kerchoffs EJH & Snorek M (Editors)*, Prague, Czech Republic, 1044-1048.

Allen, G., and Le Marshall, J. F., 1994, "An evaluation of neural networks and discrimination analysis methods for application in operational rain forecasting", *Australian Meteorological Magazine*, **43**, 17-28.

Bhattacharya, R.K., 2004, "Optimal design of unit hydrographs using probability distribution and genetic algorithms", *Sadhana*, **29**, 499-508

Baum, B.A., Tovinkere, V., Titlow, J. and Welch, R.M., 1997, "Automated cloud classification of global AVHRR data using a fuzzy logic approach", *Journal of Applied Meteorology*, **36**, 1519-1540.

Bankert, R. L. and Aha, D. W., 1996, "Improvement to a neural network cloud classifier", *Journal of Applied Meteorology*, **35**, 2036-2039

Bruton, J.M., McClendon, R.W., Hoogenboom, G., 2000, "Estimating daily Pan Evaporation with Artificial neural networks", *Trans. ASAE*, **43**, 491-496.

Bardossy, A., Duckstein, L. and Bogardi, I., 1995, "Fuzzy rule-based classification of atmospheric circulation patterns", *International Journal of Climatology*, **15**, 1087-1097

Bianco, L., and Wilczak, J.M., 2002, "Convective Boundary Layer Depth: Improved Measurement by Doppler Radar Wind Profiler Using Fuzzy Logic Methods", *Journal of Atmospheric and Oceanic Technology*, **19**, 1745-1758, doi: 10.1175/1520-0426.

Cartwright, H.M. and Harris, S.P., 1993, "Analysis of the distribution of airborne pollution using Genetic Algorithm", *Atmospheric Environment*, **27A**, 1783-1797

Chaloulakou A, Grivas G, and Spyrellis N, 2003, "Neural network and multiple regression models for PM10 prediction in Athens: a comparative assessment", *Journal of Air Waste Management Association*, **53**,183-1190.

Cook, D.F. and Wolfe, M.L., 1991, "A back-propagation neural network to predict average air temperatures", *AI Applications*, **5**, 40-46

Chaudhuri, S. and Chattopadhyay, S., 2005a, "Neuro-Computing Based Short Range Prediction of Some Meteorological Parameters during Pre-monsoon Season", *Soft Computing-A Fusion of Foundations, Methodologies and Applications*, **9**,5, 349-354, DOI 10.1007/s00500-004-0414-3

Chaudhuri, S. and Chattopadhyay, S., 2005b, "Prediction of Severe Thunderstorm with Minimal A-Priori Knowledge", *Advances in Complex Systems*, **8**, 1,75-86, DOI 10.1142/S0219525905000348

Fujibe, F., 1989, "Short-term precipitation patterns in central Honshu, Japan--- classification with the fuzzy c-means method", *Journal of Meteorological Society of Japan*, **67**, 967-982

Gardner, M.W., and Dorling, S.R., 1998, "Artificial Neural Network (Multilayer Perceptron)- a review of applications in atmospheric sciences", *Atmospheric Environment*, **32**, 2627-2636.

Gomez, V. and Casanovas, A., 2002, "Fuzzy logic and meteorological variables: a case study of solar irradiance", *Fuzzy Sets and Systems*, **126**, 121-128.

Gautam, N.K., and Kaushika, N.D., 2002, "A Model for the Estimation of Global Solar Radiation Using Fuzzy Random Variables", *Journal of Applied Meteorology*, **41**, 1267–1276, doi: 10.1175/1520-0450

Handerson, H.W., and Wells, R., 1988, "Obtaining attractor dimension for meteorological time series", *Advances in Geophysics*, **30**, 205-237

Hu, M.J.C., 1964, *Application of ADALINE system to weather forecasting*, Technical Report, Stanford Electron.

Han, H. and Felker, P., 1997, "Estimation of daily soil water evaporation using an artificial neural network", *Journal of Arid Environments*, **37**, 251-260

Hsieh, W. W. and Tang, T, 1998, "Applying Neural Network Models to Prediction and Data Analysis in Meteorology and Oceanography", *Bulletin of the American Meteorological Society*, **79**, 1855-1869.

Hansen, B.K., 2003, "Fuzzy case based prediction of cloud ceiling and visibility", Available over Internet:http://www.bjarne.ca/ams2003.pdf, Links: http://chebucto.ca/~bjarne/wind

Haupt, S.E. ,2003, "Genetic Algorithms in Geophysical Fluid Dynamics", *AMS Conference on Artificial Intelligence*, Paper P1.7.

Jagadesh, A., 2000, "Comparison of ANN and other empirical approaches for predicting watershed runoff", *Journal of Water Resources Planning and Management*, **126**, 156-166

Jef, H., Clemens, M., Gerwin, D., Frans, F. and Olivier, B., 2005, "A neural network forecast for daily average PM10 concentration in Belgium", *Atmospheric Environment*, **39**, 3279-3289

Kishtawal, C.M., Basu, S., Patadia, F. and Thapliyal, P.K., 2003, "Forecasting summer rainfall over India using Genetic Algorithm", *Geophysical Research Letters*, **30**, doi: 10.1029/2003GL018504.

Kozma,R., Kasabov, N.K., Kim, J.S. and Cohen, A., 1998, "Integration of Connectionist Methods and Chaotic Time-Series Analysis for the Prediction of Process Data", *International Journal of Intelligent Systems*, **13**, 519-538.

Kishtawal, C.M., Patadia, F, Sing, R., Basu, S., Narayanan, M.S. and Joshi, P.C., 2005, "Automatic estimation of tropical cyclone intensity using multi-channel TMI data: A genetic algorithm approach", *Geophysical Research Letters*, **32**, doi:10.1029/2004GL022045.

Lee, S., Cho, S. and Wong, P.M., 1998, "Rainfall prediction using Artificial Neural Network", *Journal of Geographic Information and Decision Analysis*, **2**, 254-264

Lakshmanan, V. and Witt, A, 1997, "A fuzzy logic approach to detecting severe updrafts", *A.I. Applications*, **11**, 1-12

Liu, H., and Chandrasekar, V., 2000, "Classification of Hydrometeors Based on Polarimetric Radar Measurements: Development of Fuzzy Logic and Neuro-Fuzzy Systems, and In Situ Verification", *Journal of Atmospheric and Oceanic Technology*, **17**, 140-164, doi: 10.1175/1520-0426

Lakshmanan, V., 2000, "Using a Genetic Algorithm to Tune a Bounded Weak Echo Region Detection Algorithm", *Journal of Applied Meteorology*, **39**, 222–230, doi: 10.1175/1520-0450

Mulligan, A. E. and Brown, L.C., 1998, "Genetic Algorithms for calibrating water quality models", *J. of Environmental Engineering*, **2**, 202-211

Maqsood, I. Muhammad, R. K., and Abraham, A., 2002, "Neurocomputing Based Canadian Weather Analysis", *Computational Intelligence and Applications*, *Dynamic Publishers Inc., USA*, 39-44.

Mohandes, A.M., S. Rehman, and T.O. Halawani, (1998), "A neural network approach for wind prediction", *Renewable Energy*, **13**, 345-354

McCann, D.W., 1992, "A Neural Network Short-Term Forecast of Significant Thunderstorms", *Weather and Forecasting*, **7**, 525-534, doi: 10.1175/1520-0434

Özelkan, E.C., Ni, F., and Duckstein, L., 1996, "Relationship between monthly atmospheric circulation patterns and precipitation: fuzzy logic and regression approaches", *Water Resources Research*, **32**, 2097-2103.

Pesti, G., Shrestha, B.P., Duckstein, L. and Bogardi, I., 1996, "A fuzzy rule-based approach to drought assessment", *Water Resources Research*, **32**, 1741-1747.

Perez, P. and Reyes, J., 2001, " Prediction of particulate air pollution using neural techniques", *Neural Computing and Application*, **10**,165-171

Perez, P., Trier, A. and Reyes, J., 2000, "Prediction of PM$_{2.5}$ concentrations several hours in advance using neural networks in Santiago, Chile", *Atmospheric Environment*, **34**, 1189-1196

Principle, J.C., Rathie, A., and Kuo, J.M., 1992, "Prediction of chaotic time series with neural networks and the issue of dynamic modeling,'' *International Journal of Bifurcation Chaos*, **2**, 989-996.

Sharifi, M. B., Georgakakos, K.P., and Rodriguez-Iturbe, R, 1990, "Evidence of deterministic chaos in the pulse of storm rainfall", *Journal of the Atmospheric Sciences*, **47**, 888-893

Sivakumar, B., Liong, S.Y., and Liaw, C.Y., 1998, "Evidence of chaotic behaviour in Singapore rainfall", *J. of American Water Resources Association*, **34**, 301-310

Stehlik, J., 1999, "Deterministic chaos in runoff series", *J. of Hydrology and Hydrodynamics*, **47**, 271-287

Sivakumar, B., 2000, "Chaos theory in hydrology: important issues and interpretations", *Journal of Hydrology*, **227**, 1-20

Shao, J., 2000, "Fuzzy Categorization of Weather Conditions for Thermal Mapping", *Journal of Applied Meteorology*, **39**, 1784-1790, doi: 10.1175/1520-0450

Sen, Z. and Oztopal, A., 2001, " Genetic algorithms for the classification and prediction of precipitation occurrence", *Hydrological Sciences*, **46**, 255-268.

Vivekanandan, J., Ellis, S.M., Oye, R., Zrnic, D.S., Ryzhkov, A.V., and Straka, J., 1999, "Cloud Microphysics Retrieval Using S-band Dual-Polarization Radar Measurements", *Bulletin of the American Meteorological Society*, **80**, 381–388, doi: 10.1175/1520-0477

Wong, K.W., Wong, P.M., Gedeon, T.D. and Fung, C.C., 1999, "Rainfall prediction using neural fuzzy technique", URL: www.it.murdoch.edu.au/~wong/publications/SIC97.pdf, 213-221

Yip, C. L. and Wong, K.Y., 2004, "Efficient and effective tropical cyclone eye fix using genetic algorithm", In: Proceedings of the 8th International Conference on Knowledge-Based Intelligent Information and Engineering Systems.

Zhang, M. and Scofield, A. R., 1994, "Artificial Neural Network techniques for estimating rainfall and recognizing cloud merger from satellite data", *International Journal of Remote Sensing*, **16**,3241-3262

Zhang, J. and Knoll, A., 2001, "Neuro Fuzzy Modeling of Time Series", In: *Soft Computing for Risk Evaluation and Management, Physica-Verlag, Heidelberg, New York*, 140-154.